\documentclass[twocolumn,secnumarabic,nobibnotes,reprint,superscriptaddress,amssymb,amsmath,aps,showpacs,10pt,floatfix,prb,longbibliography]{revtex4-2}

\usepackage{ulem}
\usepackage{graphicx}
\usepackage{amsmath}
\usepackage{amsfonts}
\usepackage{stmaryrd}
\usepackage{bbm}
\usepackage{mathrsfs}
\usepackage{tipa}
\usepackage{amssymb}
\usepackage{txfonts}
\usepackage{dcolumn}
\usepackage{epstopdf}
\usepackage{float}
\usepackage{chngcntr}
\usepackage{array}
\usepackage{longtable}
\usepackage{subfigure}
\usepackage{url}
\usepackage[utf8]{inputenc}
\usepackage{newunicodechar}
\usepackage{float}      
\usepackage{titlesec}
\usepackage[colorlinks,bookmarks=true,citecolor=blue,linkcolor=blue,urlcolor=blue]{hyperref} 

\begin{document}

\title{Hidden phonon-assisted charge density wave transition in BaFe$_2$Al$_9$\\
	revealed by ultrafast optical spectroscopy}

\author{Lei Wang}
\affiliation{Center for Advanced Quantum Studies and Department of Physics, Beijing Normal University, Beijing 100875, China}
\affiliation{Beijing National Laboratory for Condensed Matter Physics, Institute of Physics, Chinese Academy of Sciences, Beijing 100190, China}

\author{Mingwei Ma}
\affiliation{Beijing National Laboratory for Condensed Matter Physics, Institute of Physics, Chinese Academy of Sciences, Beijing 100190, China}
\affiliation{School of Physical Sciences, University of Chinese Academy of Sciences, Beijing 100190, China}

\author{Jiangxu Li}
\affiliation{Shenyang National Laboratory for Materials Science, Institute of Metal Research, Chinese Academy of Sciences, 110016 Shenyang, China}

\author{Liucheng Chen}
\affiliation{Beijing National Laboratory for Condensed Matter Physics, Institute of Physics, Chinese Academy of Sciences, Beijing 100190, China}
\affiliation{School of Physical Sciences, University of Chinese Academy of Sciences, Beijing 100190, China}

\author{Bingru Lu}
\affiliation{Beijing National Laboratory for Condensed Matter Physics, Institute of Physics, Chinese Academy of Sciences, Beijing 100190, China}

\author{Xiang Li}
\affiliation{Beijing National Laboratory for Condensed Matter Physics, Institute of Physics, Chinese Academy of Sciences, Beijing 100190, China}

\author{Feng Jin}
\affiliation{Beijing National Laboratory for Condensed Matter Physics, Institute of Physics, Chinese Academy of Sciences, Beijing 100190, China}

\author{Elbert E. M. Chia}
\affiliation{Division of Physics and Applied Physics, School of Physical and Mathematical Sciences, Nanyang Technological University, Singapore 637371, Singapore}

\author{Jianlin Luo}
\affiliation{Beijing National Laboratory for Condensed Matter Physics, Institute of Physics, Chinese Academy of Sciences, Beijing 100190, China}
\affiliation{School of Physical Sciences, University of Chinese Academy of Sciences, Beijing 100190, China}

\author{Rongyan Chen}
\email{rychen@bnu.edu.cn}
\affiliation{Center for Advanced Quantum Studies and Department of Physics, Beijing Normal University, Beijing 100875, China}

\author{Peitao Liu}
\email{ptliu@imr.ac.cn}
\affiliation{Shenyang National Laboratory for Materials Science, Institute of Metal Research, Chinese Academy of Sciences, 110016 Shenyang, China}

\author{Fang Hong}
\email{hongfang@iphy.ac.cn}
\affiliation{Beijing National Laboratory for Condensed Matter Physics, Institute of Physics, Chinese Academy of Sciences, Beijing 100190, China}
\affiliation{School of Physical Sciences, University of Chinese Academy of Sciences, Beijing 100190, China}

\author{Xinbo Wang}
\email{xinbowang@iphy.ac.cn}
\affiliation{Beijing National Laboratory for Condensed Matter Physics, Institute of Physics, Chinese Academy of Sciences, Beijing 100190, China}

\date{\today}

\begin{abstract}

The interplay between electronic and lattice degrees of freedom is fundamental to charge density wave (CDW) formation, yet the microscopic origin often remains elusive. Here, we investigate the transient optical response of the intermetallic compound BaFe$_2$Al$_9$ using polarization-resolved ultrafast optical spectroscopy. We identify a discontinuous sign reversal in the transient reflectivity at T$_C$$\simeq$ 110 K, providing unambiguous evidence for the first-order transition. The anisotropic quasiparticle relaxation establishes the three-dimensional nature of the ordered state. Below T$_C$, a single coherent 1.6 THz oscillation appears abruptly and remains confined to the CDW phase. This mode exhibits weak temperature dependence with negligible softening and is absent in Raman spectra. First-principles calculations imply that it is a precursor phonon at the CDW wave vector with strong electron-phonon coupling. Our results indicate that the CDW in BaFe$_2$Al$_9$ arises from intertwined electronic and lattice instabilities, assisted by a displacive mechanism mediated by a hidden strongly coupled phonon---distinct from conventional amplitude-mode softening scenarios.

\end{abstract}

\maketitle

\section{Introduction}

Charge density wave is a prototypical broken-symmetry ground state in correlated electron systems, characterized by a periodic modulation of the electron density and a concomitant periodic lattice distortion. The conventional Peierls mechanism, established for low-dimensional systems, attributes CDW formation to Fermi-surface nesting at a characteristic wave vector $q_{\rm CDW}$. This electronic instability is expected to induce a Kohn anomaly, a sharp softening of a phonon mode that ultimately condenses into the static lattice distortion below the transition temperature  \cite{frohlich1954, peierls1955, kohn1959}. However, in higher-dimensional materials this nesting-only picture is often insufficient for stabilizing the CDW order  \cite{cdw1, cdw2, cdw6, TiTe2}. Instead, the instability can be understood to emerge from a cooperative interplay between a favorable electronic susceptibility and strong electron-phonon coupling (EPC)   \cite{cdw3, cdw4, cdw5}. This interplay becomes particularly intricate in three-dimensional systems, where the precise microscopic drivers often remain elusive, motivating further investigation into novel CDW materials.

The intermetallic compound  BaFe$_2$Al$_9$ crystallizes in a hexagonal P6/mmm structure, where the Al atoms form a kagome lattice while the Fe atoms form a honeycomb lattice centered by the Ba atoms. It undergoes a pronounced first-order structural transition near 100 K, evidenced by sharp anomalies and thermal hysteresis in thermodynamic and transport properties   \cite{catastrophic, Thermal}. Below T$_C$, the emergence of superlattice peaks in X-ray and neutron diffraction reveals a complex three-dimensional CDW state   \cite{catastrophic, 27AL}. Structural refinements indicate a primary modulation of Fe chains along the c axis, stabilized by cooperative Ba displacements, yet the microscopic origin of the transition remains unsettled. Early first-principles study invoked Fermi-surface nesting at a characteristic wave vector involving Fe-3$d$ states \cite{Origin}, whereas $^{57}$Fe M\"ossbauer spectroscopy resolves two inequivalent Fe environments, implying a charge distribution more complex than a simple sinusoidal modulation \cite{catastrophic}. Strikingly, recent high-pressure measurements report an anomalous enhancement of T$_C$ to room temperature around 3 GPa, opposite to the suppression expected for conventional nesting-driven CDW systems  \cite{HP1, HP2}.  These observations suggest that the simple nesting scenario is incomplete and that other factors, such as strong EPC, may play a decisive role. To date, experimental work has largely focused on static structural and transport properties, leaving the dynamical interplay between the electronic structure and the lattice across the transition unresolved. Therefore, a direct probe of the coupled electron-phonon dynamics is essential to elucidate the true driving mechanism of the CDW transition in BaFe$_2$Al$_9$.

Time-resolved pump-probe spectroscopy is widely used to investigate the nonequilibrium dynamics in CDW materials, owing to its high sensitivity to the opening of small energy gap and low-energy collective excitations  \cite{KMoO3,Elbert07,Ba12210,Meng327,Giannetti03032016,dong2023recent}. Polarization-resolved measurements can further resolve anisotropic transient responses and coupling pathways. For instance, polarization-resolved studies have disentangled the electronic and lattice contributions to the order parameter in ZrTe$_3$  \cite{ZrTe3}, and have uncovered a hidden three-dimensional component of the CDW order in CuTe  \cite{3DCuTe}. Given this demonstrated capability to reveal anisotropic interactions and emergent ordering, the technique is particularly well suited to investigate the CDW state in BaFe$_2$Al$_9$.

Here, we employ polarization-resolved ultrafast optical spectroscopy to track the non-equilibrium dynamics across the CDW transition in BaFe$_2$Al$_9$. We observe a discontinuous sign reversal in the transient reflectivity around 110 K, providing clear evidence for a first-order transition with pronounced three-dimensional anisotropy in the quasiparticle response. Within the CDW phase, a single coherent 1.6 THz oscillation appears abruptly, exhibits negligible thermal softening and is absent in Raman spectra. In conjunction with first-principles calculations, we ascribe this oscillation to a Ba-dominated phonon at $q_{\rm CDW}$ with strong EPC which becomes coherently activated only below T$_C$ via a displacive mechanism. Our results indicate that the CDW in BaFe$_2$Al$_9$ arise from intertwined electronic and lattice instabilities, with a hidden strongly coupled phonon mediating the coherent dynamics beyond the standard amplitude-mode softening paradigm.

\section{Experimental and computational methods}

Single crystals of BaFe$_2$Al$_9$ were synthesized using the Al self-flux method, following procedures detailed in Ref. \cite{catastrophic}. For optical measurements, a small shiny fragment with $\sim$200 $\mu$m lateral size were selected to minimize stress-induced fragmentation across the structural transition. The crystallographic orientation of the sample was determined using a single-crystal X-ray diffractometer equipped with a rotating Ag-anode source.  The ultrafast pump-probe experiments were performed using an optical parametric amplifier seeded by a 50 kHz Yb:KGW amplifier, generating 800 nm pulses with a duration of 50 fs. The output beam was split into pump and probe arms. The pump beam was frequency-doubled to 400 nm with BBO crystal.  Both beams were aligned collinearly and focused normal to the sample surface with a 5$\times$ microscope objective, resulting in spot diameters of 34 and 14 $\mu$m for pump and probe beams, respectively. The measurements were performed at a fixed pump fluence of 45 $\mu J/cm^2$. The linear polarization of both beams were controlled using a half-wave plate and a wire-grid polarizer. Samples were mounted in a continuous-flow helium cryostat, and measurements were performed upon warming from the base temperature. 

First-principles calculations were performed using the the QUANTUM ESPRESSO package~\cite{Giannozzi_2009,Giannozzi_2017}. Norm-conserving Vanderbilt pseudopotentials~\cite{PhysRevB.88.085117} with an energy cutoff of 55 Ry were used. The structures were relaxed until the energy and force were smaller than 10$^{-9}$ Ry and 10$^{-7}$ Ry/bohr, respectively. The EPC calculations were conducted using the EPW code~\cite{PONCE2016116}. The Wannier functions were constructed using the Ba-$s$, Al-$sp$, and Fe-$d$ states~\cite{Origin}.
The mode-resolved EPC is calculated as
\begin{equation}
	\lambda_{\mathbf{q}\nu} = \frac{2 S(\mathbf{q}\nu)}{N(E_F) \omega_{\mathbf{q}\nu}},
	\label{eq:mode_EPC}
\end{equation}
where $\omega_{\mathbf{q}\nu}$ is the $\nu$-branch phonon frequency at wave vector $\mathbf{q}$,
$N(E_F)$ represents the electronic density of states at the Fermi energy $E_F$,
and $S(\mathbf{q}\nu)$ is calculated by
\begin{equation}
	S(\mathbf{q}\nu) \equiv \frac{1}{N_k}
	\sum_{mn\mathbf{k}}
	|g_{mn\nu}(\mathbf{k},\mathbf{q})|^2
	\delta(\varepsilon_{n\mathbf{k}} - E_F)
	\delta(\varepsilon_{m\mathbf{k}+\mathbf{q}} - E_F) .
	\label{eq:Sqv}
\end{equation}
Here, $g_{mn\nu}(\mathbf{k},\mathbf{q})$ is the EPC matrix element and $\varepsilon_{n\mathbf{k}}$ denotes the Kohn-Sham one-electron energy.
With Eq.~\eqref{eq:mode_EPC}, the Eliashberg spectral function is calculated as
\begin{equation}
	\begin{split}
		\alpha^2F(\omega) = \frac{1}{2} \sum_{\mathbf{q}\nu}
		\frac{\lambda_{\mathbf{q}\nu}}{\omega_{\mathbf{q}\nu}}
		\delta(\omega - \omega_{\mathbf{q}\nu}).
	\end{split}
	\label{eq:a2f_lambda}
\end{equation}
Then, the cumulative EPC parameter can be calculated by
\begin{equation}
	\lambda(\omega) = 2 \int_0^{\omega} \frac{\alpha^2F(\omega')}{\omega'} \, d\omega',
	\label{eq:lambda_def}
\end{equation}
and the EPC constant is computed as
\begin{equation}
	\lambda \equiv \lambda(\infty) = 2 \int_0^\infty \frac{\alpha^2F(\omega')}{\omega'} \, d\omega'.
\end{equation}
The derivative of $\lambda(\omega)$ yields the marginal contribution to $\lambda$,
\begin{equation}
	\frac{d\lambda(\omega)}{d\omega} = \sum_{\mathbf{q}\nu} \lambda_{\mathbf{q}\nu} \, \delta(\omega - \omega_{\mathbf{q}\nu}).
	\label{eq:lambda_diff}
\end{equation}

\section{Results and discussion}

\begin{figure}[tpb]
\centering
\includegraphics[width=0.95\columnwidth]{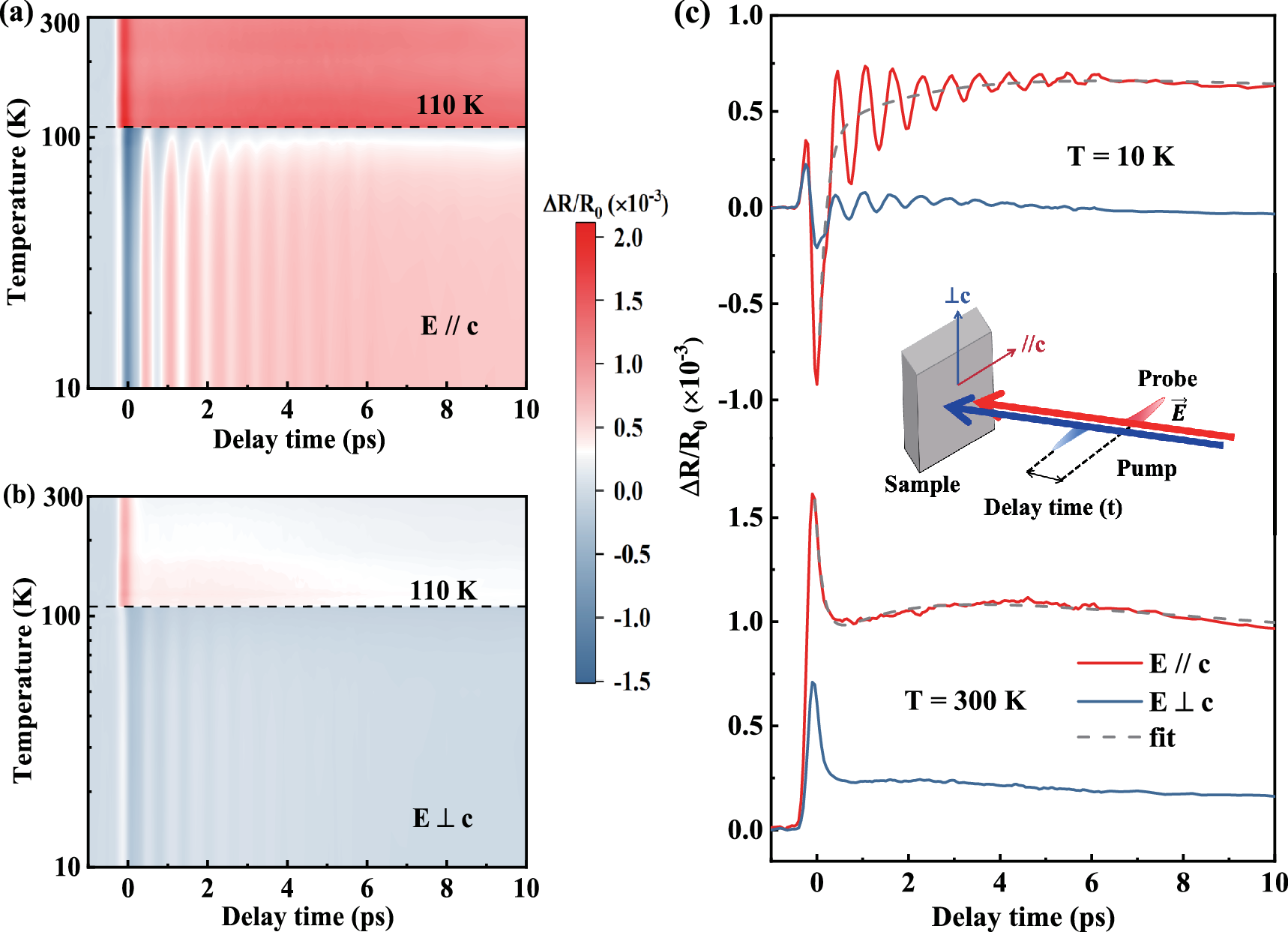}
\subfigure{\label{1a}}
\subfigure{\label{1b}}
\subfigure{\label{1c}}
\caption{Temperature-dependent ultrafast dynamics for BaFe$_2$Al$_9$. Two-dimensional color maps of the transient reflectivity change, $\Delta R/R$, as a function of pump-probe time delay and temperature for (a) E\,//\,c and (b) E\,$\perp$\,c. The horizontal dashed lines mark the CDW transition temperature. (c) Representative transients at 10 K (upper) and 300 K (lower) for both polarizations up to 10 ps. Dash curves are the fits to the three exponential relaxation background. Inset: schematic of the polarization-resolved pump-probe experiment.} 
\label{Fig1}
\end{figure}

Figures \ref{1a} and \ref{1b} present the temperature evolution of the transient reflectivity change $\Delta R(t)/R_0$, for pump and probe polarization aligned parallel (E\,//\,c) and perpendicular (E\,$\perp$\,c) to the crystallographic c-axis, respectively. The transient response exhibits a similar trend for both polarization configurations, featuring an abrupt and discontinuous change upon warming through T$_C$ $\simeq$110 K. In particular, the sign of the initial reflectivity change reverses from negative to positive above the critical temperature. Furthermore, superimposed on the relaxation dynamics, coherent oscillations are clearly resolved in the low-temperature phase, which abruptly vanish across T$_C$. These pronounced changes in the transient optical response are a clear signature of the first-order structural transition in BaFe$_2$Al$_9$, which occurs near 100 K and exhibits a thermal hysteresis of approximately 10 K between heating and cooling. This phase transition is associated with the CDW transition \cite{catastrophic,27AL,Thermal}.  

To illustrate the polarization-dependent quasiparticle dynamics, Fig. \ref{1c} compares the transient reflectivity measured with E\,//\,c and E\,$\perp$\,c at two representative temperatures, 10 K and 300 K. The response exhibits a strong optical anisotropy, with the magnitude of the transient signal being consistently larger for the E\,//\,c configuration. Notably, despite this large anisotropy in the relaxation dynamics, a small initial spike observed near zero time delay below T$_C$ maintains a nearly constant amplitude for both polarizations. This feature is likely a coherent artifact arising from the pump-probe correlation  \cite{coherentartifact1,coherentartifact2,coherentartifact3}. Therefore, the following  analysis focuses on the relaxation dynamics after this initial artifact. To quantitatively analyze the carrier relaxation dynamics, the non-oscillatory component of $\Delta R/R$ signal was fitted by a sum of three-exponential decay function, ${\Delta R(t)}/{R_0} = \sum_{i=1}^{3} A_i \exp(-t/\tau_i) + C$, where $A_i$ and $\tau_{i}$ are the amplitude and relaxation time of the \textit{i}-th component, respectively, and C is a constant offset accounting for the thermal effects at long delay times. Representative fits at 10 and 300 K for E\,//\,c polarization are shown as the dash lines in Fig. \ref{1c}, demonstrating good agreement between experimental data and the fitted curves.

\begin{figure}[tpb]
\centering
\includegraphics[width=0.95\columnwidth]{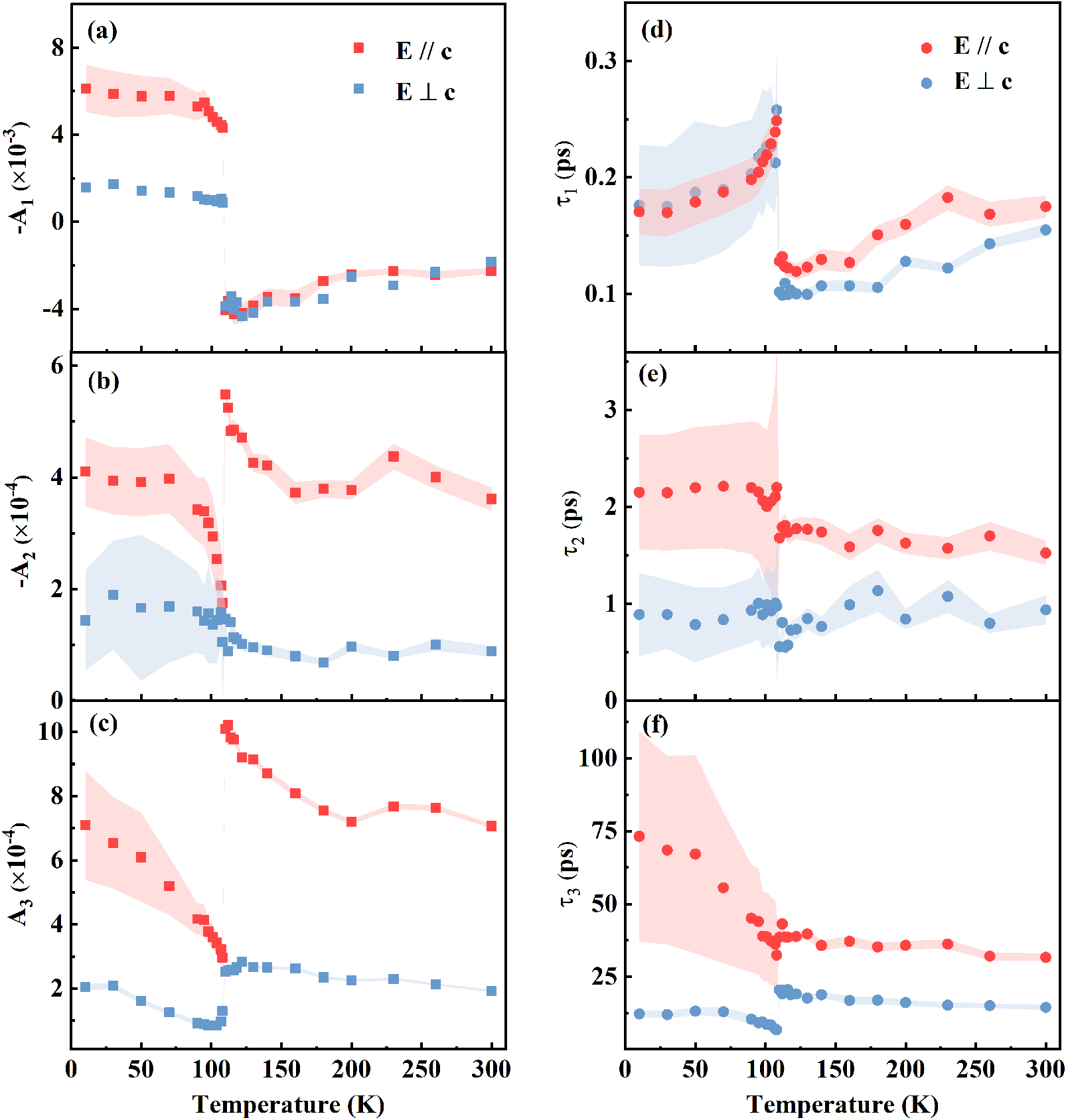}
\vspace{-0.5cm}
\subfigure{\label{2a}}
\subfigure{\label{2b}}
\subfigure{\label{2c}}
\subfigure{\label{2d}}
\subfigure{\label{2e}}
\subfigure{\label{2f}}
\caption{Quantitative analysis of the quasiparticle relaxation. Temperature dependence of the (a-c) amplitudes and (d-f) relaxation times extracted from the three exponential fit to the experimental data. Error bands indicate fit uncertainties.}
\label{Fig2}
\end{figure}

A quantitative analysis of the extracted parameters, as depicted in Fig. \ref{Fig2}, reveals a clear restructuring of quasiparticle dynamics across the CDW transition. Above T$_C$, $A_1$ amplitude increases gradually while $\tau_{1}$ decreases with cooling, consistent with metallic behavior in BaFe$_2$Al$_9$. Upon entering the CDW phase, $A_1$ reverses sign from positive to negative, indicating an abrupt electronic reconstruction. Meanwhile, $\tau_{1}$ grows sharply as a partial gap opens on the Fermi surface, reducing the phase space for recombination  \cite{catastrophic, Thermal}. Approaching to T$_C$,  $\tau_{1}$ exhibits a pronounced upturn for both polarizations while $A_1$ amplitude decreases monotonically, consistent with a phonon-bottleneck effect governed by a temperature-dependent gap that narrows near T$_C$  \cite{RT}. In addition, the slower components likewise show step-like changes across the transition. The concerted discontinuous evolution of all channels provides compelling evidence for the first-order nature of the coupled CDW and structural transitions in BaFe$_2$Al$_9$, in line with transient-optical fingerprints reported in diverse systems, including the kagome superconductor CsV$_3$Sb$_5$  \cite{CsVSb,CsVSbPRM}, the quasi-1D system (TaSe$_4$)$_2$I  \cite{Schaefer}, the non-magnetic BaNi$_2$As$_2$  \cite{BaNiAs}, and pressurized topological insulator Bi$_2$Se$_3$   \cite{BiSe}.

\begin{figure}[htpb]
\centering
\includegraphics[width=0.95\columnwidth]{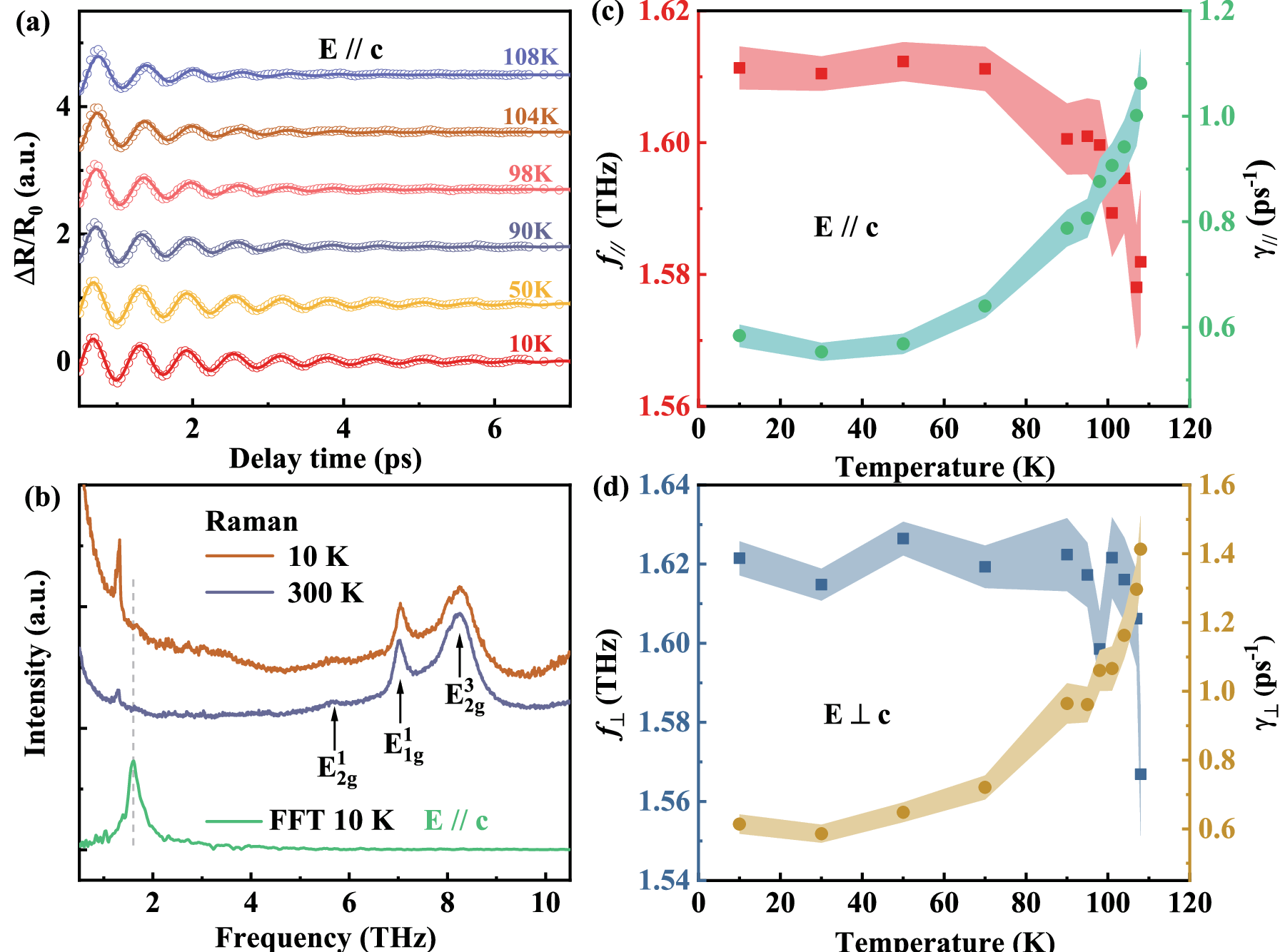}
\subfigure{\label{3a}}
\subfigure{\label{3b}}
\subfigure{\label{3c}}
\subfigure{\label{3d}}
\caption{Coherent phonon spectroscopy in the CDW phase. (a) Oscillatory component of the transient reflectivity for E\,//\,c at various temperatures below T$_C$, isolated by subtracting the multi-exponential background. (b) Comparison of the FFT spectrum of the oscillatory signal at 10 K with experimental Raman spectra. Arrows mark the observed Raman-active phonon modes, with their assignments guided by first-principles calculations. Dash line highlights the dominant 1.6 THz mode, absent in the Raman spectra. (c-d) Temperature dependence of the oscillation frequency $f$ and damping rate $\gamma$ for both polarizations. Error bands indicate fit uncertainties.}\label{Fig3}
\end{figure}

\begin{figure*}[tpb]
	\centering
	\includegraphics[width=1.65\columnwidth]{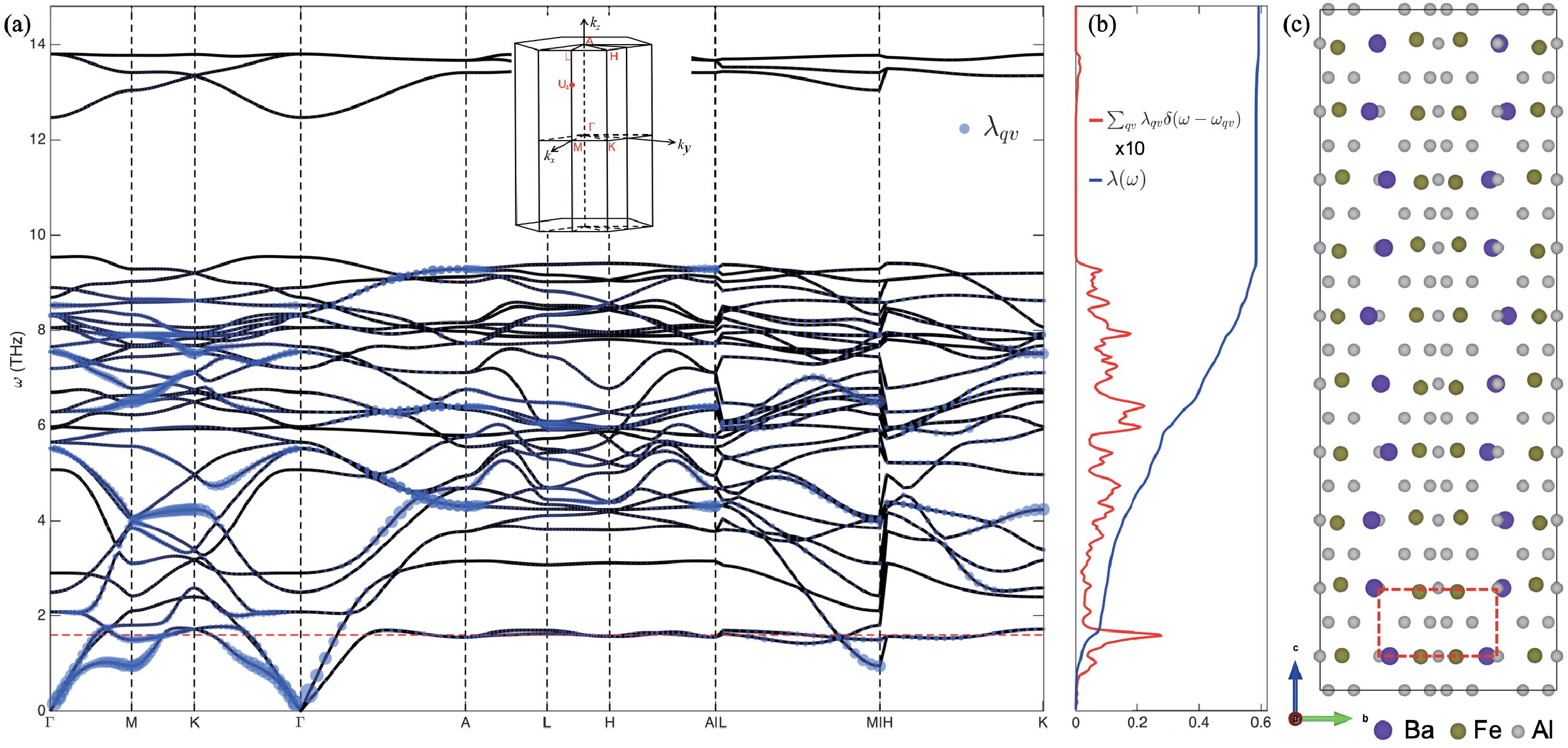}
	\vspace{-0.2cm}
	\subfigure{\label{4a}}
	\subfigure{\label{4b}}
	\subfigure{\label{4c}}
	\subfigure{\label{4d}}
	\caption{First-principles analysis of the phonon dispersion and EPC for BaFe$_2$Al$_9$ in the high temperature P6/mmm phase. (a) Phonon dispersion superimposed by the mode-resolved EPC (blue circles, Eq.~\eqref{eq:mode_EPC}), with radius proportional to the coupling magnitude. A horizontal dashed line highlights the 1.6 THz frequency. Inset: The Brillouin zone with high symmetry points. The U$_4$ point lies on the L-M line at fractional coordinates (0.5, 0, 0.3). (b) Cumulative EPC parameter (blue curve, Eq.~\eqref{eq:lambda_def}) and  the marginal contribution to $\lambda$ (red curve, Eq.~\eqref{eq:lambda_diff}), magnified by 10 times for clarify. (c) Real-space visualization of the atomic displacements for the 1.6 THz phonon mode at the U$_4$ point, highlighting the dominant sine-modulated in-plane vibrations of the Ba atoms. The dashed rectangle denotes the unit cell of the high-temperature phase.}\label{Fig4}
\end{figure*}

Beyond the quasiparticle relaxation, the transient reflectivity traces display a pronounced coherent oscillation superimposed on the slowly decaying background that appears exclusively within the CDW phase. After subtracting the multi-exponential background, the residual is dominated by a single well-defined oscillation [Fig.  \ref{3a}], while the Fast Fourier Transform (FFT) spectrum shows a sharp peak at approximately 1.6 THz  [Fig.  \ref{3b}]. In general, coherent phonons manifest as real-time oscillations in pump-probe measurements and are typically generated via either the impulsive stimulated Raman scattering (ISRS) or the displacive excitation of coherent phonons (DECP) mechanism  \cite{Merlin1997}. Given the proximity of this mode to the first-order structural transition, the oscillation may originate from a coherent lattice vibration that becomes allowed only in the symmetry-broken phase. To test this hypothesis, we performed temperature-dependent Raman-scattering measurements. As shown in Fig.  \ref{3b}, no Raman-active phonon is observed near 1.6 THz in either the high-temperature phase or the low-temperature CDW phase. We note that the narrow feature near 1.3 THz is likely an experimental artifact (e.g., atmospheric scattering) and is not related to intrinsic lattice vibrations. Three higher-frequency modes, located at 5.7, 7 and 8.2 THz, are clearly resolved and exhibit negligible temperature dependence. Their frequencies are in good agreement with the Raman-active E$_g$ modes predicted by our first-principles calculations (see Table \ref{table1} and discussion below). These results provide decisive evidence that the 1.6 THz mode can not be attributed to a conventional optical phonon.

To perform a quantitative analysis of the coherent dynamics, the oscillatory component was fitted using a damped cosine function, $\Delta R / R_{osc}(t) = B e^{-\gamma t} \cos(2\pi f t + \phi)$, where \( B, \gamma, f, \) and \( \phi\) are the oscillatory signal amplitude, damping rate, frequency, and initial phase, respectively. Excellent fits, shown as solid curves in Fig.  \ref{3a},were obtained for all temperatures below T$_C$. The temperature dependencies of the oscillation frequency and damping rate for both polarization geometries are shown in Figs.  \ref{3c} and \ref{3d}. Upon warming towards T$_C$, the frequency undergoes a slight but distinct softening ($\sim$2\% for E\,//\,c, even weaker for E\,$\perp$\,c), while the damping rate exhibits a pronounced increase for both polarizations. Such behavior resembles that of CDW amplitude modes---a collective excitation of the order parameter's magnitude observed in numerous systems   \cite{BiRhSe,CuTe,LaAgSb,ScV6Sn6,WeiZhengxin, CsVSb}. Nevertheless, the exceptionally weak softening observed in BaFe$_2$Al$_9$\ stands in contrast to the complete softening expected for a conventional Peierls-type CDW amplitude mode, where the frequency vanishes at T$_C$ in a continuous second-order transition   \cite{KMoO3}. 

\begin{table}[tp]
	\centering
	\caption{Raman-active phonon modes of BaFe$_2$Al$_9$ at $\Gamma$. Experimental frequencies were determined at 10 K. Theoretical values were calculated for the high-temperature phase with P6/mmm space group. Symmetry assignments correspond to the irreducible representations of the D$_{6h}$ point group.}
	\label{table1}
	\setlength{\tabcolsep}{14pt}
	\renewcommand{\arraystretch}{1.5} 
	\begin{tabular}{cccc}
		\hline\hline
 		\multicolumn{4}{c}{Space group P6/mmm} \\
		\hline
		& Symm. & Expt. (THz) & Calc. (THz) \\
		\hline
		1  & $E_{2g}^1$      &    5.7   & 5.44  \\
		2  & $E_{1g}$       & 7 &    7.1   \\
		3  & $E_{2g}^2$        & - &    7.47   \\
		4  & $E_{2g}^3$     & 8.2  & 8.31  \\
		5  & $A_{1g}^1$     & -  & 8.5  \\
		\hline\hline
		
	\end{tabular}
\end{table}

The absence of critical softening is consistent with the first-order character of the CDW transition in BaFe$_2$Al$_9$. Similar behavior has been reported in other materials exhibiting first-order CDW-related transitions, where coherent oscillations appear only below the structural transition   \cite{CsVSb, CsVSbPRM, Schaefer, BaNiAs,TaTe2}. In these systems, the oscillatory modes have been attributed either to a finite-momentum phonon from the high-temperature phase that becomes Raman-active at the Brillouin zone center via zone-folding, or to an amplitude mode whose softening is interrupted by the first-order transition. In BaFe$_2$Al$_9$, however, no corresponding phonon is observed in the low-temperature Raman spectrum as shown in Fig. \ref{3b}. Furthermore, the temperature dependence of the extracted damping rate can not be described by the anharmonic phonon-decay model   \cite{anharmonic}. Instead, its evolution is similar to prototype CDW systems, where the damping scales inversely with the order parameter   \cite{Disentanglement, Collectivemodes}. These findings rule out the zone-folding scenario and suggest that the 1.6 THz mode is not a conventional amplitude mode. We instead propose that this mode originates from a more complex coupling between the electronically driven symmetry breaking and a specific lattice vibration intrinsic to BaFe$_2$Al$_9$.

To elucidate the microscopic origin of the observed 1.6 THz mode, we performed first-principles calculations of the phonon dispersion of BaFe$_2$Al$_9$ in its high temperature P6/mmm structure. The calculated phonon dispersion, shown in Fig.  \ref{4a}, exhibits no imaginary frequencies throughout the Brillouin zone, indicating the dynamical stability of the parent phase. From the dispersion, the Raman-active modes at the $\Gamma$ point were identified and are summarized in Table \ref{table1}. No optical phonon mode is present near 1.6 THz at $\Gamma$. The calculated phonon frequencies and symmetries at the $\Gamma$ point are in good agreement with the experimentally observed Raman modes, confirming the reliability of the calculations. Moreover, the results reveal a nearly dispersionless optical branch at approximately 1.6 THz that extends across a wide region of the Brillouin zone. Analysis of the eigenvectors associated with the 1.6 THz phonon at the CDW wave vector $q_{\rm CDW}$=$U_4$=(0,5, 0, 0.3)~\cite{catastrophic}, as visualized in Fig.~\ref{4c}, reveals the dominant sine-modulated in-plane vibrations of the Ba atoms that are weakly coupled to the surrounding Fe-Al framework. The coincidence in both energy and momentum indicates that the experimentally observed 1.6 THz coherent oscillation is directly linked to the CDW transition.

To quantify the interplay between electronic states and lattice dynamics in BaFe$_2$Al$_9$, we calculated the mode-resolved EPC, demonstrating the strong coupling of the electronic states near the Fermi surface with specific phonon modes. In particular, the phonon modes around the $\Gamma$ point exhibit the strongest EPC, where the vibrations of Ba atoms predominate. Notably, the cumulative EPC parameter exhibits rapid increase at 1.6 THz (blue curve in Fig.~\ref{4b}). This leads to a sharp peak centered at 1.6 THz in the marginal contribution to $\lambda$  (red curve in Fig.~\ref{4b}), indicating the important role of the 1.6 THz phonon modes. This resonance originates from the high phonon density of states associated with the flat phonon bands around 1.6 THz, thereby enhancing the phase space for electron-phonon scattering. Within a DECP framework, photoexcited carriers rapidly thermalize and dissipate energy via the phonon branch with the largest EPC, making the 1.6 THz mode to be the primary dissipation pathway. The direct correspondence between the EPC peak and the coherent oscillation observed in ultrafast reflectivity measurements reveals that the observed mode is a real-time manifestation of the lattice vibration most strongly intertwined with low-energy electronic states.

Our calculations provide a microscopic picture for the CDW formation in BaFe$_2$Al$_9$. The strong electronic instability originating from the Fermi surface nesting predisposes the system to a periodic lattice distortion at the CDW wave vector  $q_{\rm CDW}$~\cite{Origin}. Crucially, the partially-filled Fe-$3d$ electronic states near the Fermi surface contributing to the divergent electronic susceptibility couple selectively to a special Ba-dominated 1.6 THz phonon mode with strong EPC. Such coupled electron-phonon interplay furnishes the lowest-energy channel to accommodate the instability, thereby acting as the primary driver of the CDW transition. This electronic-instability-driven, phonon-assisted scenario consistently explains our experimental observations: ($i$) negligible frequency softening, characteristic of a strong first-order transition, ($ii$) Raman inactivity, as expected for a non-zone-center precursor mode, and ($iii$) direct correspondence with the CDW wave vector. Therefore, the 1.6 THz coherent oscillation observed below T$_{c}$ is not merely a consequence of lower symmetry but the direct dynamical signature of the phonon modes linked to the CDW phase transition.

\section{Conclusion}

In summary, we have performed  ultrafast optical spectroscopy and first-principles calculations on the intermetallic compound BaFe$_2$Al$_9$ . Our measurements reveal a first-order CDW transition near 110 K, evidenced by a discontinuous change in the transient reflectivity. A pronounced anisotropy in the quasiparticle relaxation underscores the three-dimensional character of the ordered state. Below T$_C$, a coherent 1.6 THz mode emerges abruptly and remains confined to the CDW phase. Unlike a conventional soft amplitude mode, this excitation exhibits negligible frequency softening upon warming and is absent in Raman spectra. Assisted by first-principles calculations, we identify this mode as a collective amplitude mode originating from a precursor phonon at the CDW wave vector that is coherently driven only within the symmetry-broken state with strong electron-phonon coupling. Our results demonstrate that the CDW transition in BaFe$_2$Al$_9$ is not purely electronically driven but is assisted by a displacive mechanism mediated by a hidden strongly coupled phonon that becomes optically active only below the transition temperature.

\section{Acknowledgements}
This work was supported by the National Key R\&D Program of China (Grant No. 2024YFA1611300), the National Natural Science Foundation of China (Grants No.  12574349, No. 52422112 and No. 12134018). E.E.M.C. acknowledges support from the Singapore Ministry of Education (MOE) Academic Research Fund Tier 3 (MOE-MOET32023-0003) grant. This work was supported by the Synergetic Extreme Condition User Facility (SECUF, https://cstr.cn/31123.02.SECUF).

\bibliographystyle{apsrev4-2}
\bibliography{BFA_1006.bib}

\end{document}